\documentstyle[prb,preprint,aps,psfig]{revtex}

\begin{document}

\title{Aging in disordered systems}

\author{Heiko Rieger}

\address{
Institut f\"ur Theoretische Physik,
Universit\"at zu K\"oln, 50937 K\"oln, Germany\\
and HLRZ, Forschungszentrum J\"ulich, 52425 J\"ulich, Germany\\
\smallskip
{\rm e-mail: rieger@thp.uni-koeln.de}}

\date{\today}

\maketitle

\begin{abstract}

  The dynamics of strongly disordered systems becomes extremely slow
  or glassy at low temperatures, which results in a characteristic
  aging scenario. This means that the outcome of measurements strongly
  depends on the history of the system within the glassy phase, even
  on macroscopic time scale like hours, weeks or years. This area of
  non-equilibrium dynamics in disordered systems became recently a
  major focus of research interest, in particular with respect to spin
  glasses and related systems. Here we give an overview on these
  activities with a focus on Monte Carlo studies.

\end{abstract}


\baselineskip=18pt

\newcommand{\bc}{\begin{center}}
\newcommand{\ec}{\end{center}}
\newcommand{\ba}{\begin{array}}
\newcommand{\ea}{\end{array}}
\newcommand{\bt}{\begin{tabular}}
\newcommand{\et}{\end{tabular}}
\newcommand{\bi}{\begin{itemize}}
\newcommand{\ei}{\end{itemize}}
\newcommand{\be}{\begin{equation}}
\newcommand{\ee}{\end{equation}}

\section{Introduction}

The physics of strongly disordered systems has been a focal point of
research by experimentalists and theoreticians in recent years.  Among
the most studied materials/models in this category are magnetic
systems with impurities and especially spin glasses \cite{Review,MCreview}.
However, their glassy features observed in experiments as well as the
theoretical problems occurring in spin glass models are also
encountered in various different situations: So for instance in the
context of structural glasses \cite{glasses}, manifolds (e.g.\
polymers) in random media \cite{manifolds}, protein folding
\cite{protein}, dirty superconductors \cite{super},
charge--density--wave systems with impurities \cite{CDW} and also
other areas like neural networks, population dynamics, immunology and
biological evolution \cite{biology}.

Much of the theoretical effort has concentrated on the existence
and characterization of an equilibrium phase transition in
spin glass models. Here much progress has been made, although
some questions still await a final answer \cite{MCreview}.
An at least equally fascinating subject, for which theory is
even harder, is the description of its non-equilibrium dynamics.
Experimentally this is the predominant scenario one has to deal
with, since equilibration of a spin glass is nearly impossible
on laboratory time scales. A major part of glassy dynamics takes
place out of equilibrium and a characteristic feature of it is
what one calls {\it aging}. This has been first observed by
experimentalists 10 years ago \cite{Lundgren} and denotes the
striking dependence of e.g.\ magnetization measurements in spin glasses
on the history (or age) of the system within the frozen phase.

A huge number of experimental investigations of this phenomenon
has been performed since then \cite{Ledermann,Vincent,Mattson} and it turned
out that
it occurs quite commonly in all different kinds of spin glass materials
and can also be observed in other disordered substances like in
amorphous polymers \cite{Struik}, high--$T_c$ superconductors
\cite{Rossel} or in charge-density-wave systems \cite{Biljakovic}.
Furthermore it does not depend on the existence of an
equilibrium phase transition, for instance also two-dimensional spin glasses
exhibit aging \cite{exp2d}. We will argue below that it is a
characteristic feature of any disordered or amorphous system with
a broad distribution of relaxation times and is therefore
observable also in the various contexts mentioned above.

On the theoretical side the aging phenomenon in spin glasses has
attracted ever increasing interest in recent years. Due to its extreme
mathematical difficulties it was not possible up to now to formulate a
microscopic theory for the non-equilibrium dynamics of finite
dimensional spin glasses. Only very recently progress has been made in
the analytic treatment of aging in mean-field models of Ising spin
glasses \cite{skage} and simplified or toy models \cite{Bouchaud}.
However, it has not been possible yet to make quantitative predictions
that would render a comparison with experiments possible. More
successful in this respect are phenomenological theories like the
droplet model by Fisher and Huse \cite{FiHu} and the domain theory by
Koper and Hilhorst \cite{KoHi}.

Therefore numerical studies play a decisive role in discriminating
between various predictions of the existing phenomenological models.
As is well known in a Monte-Carlo simulation one can obtain much more
detailed information about the dynamical processes and the spatial
correlations in the system under consideration. Moreover with
increasing computer power one might even be able to come close to the
macroscopic time scales that are relevant in real experiments. Much
progress has been obtained recently and in this paper we intend to
give an overview on our recent numerical studies on aging in spin
glasses and other strongly disordered systems.

\section{Aging in spin glasses}

The first observation that one makes by dealing with spin glasses and
related systems is that their dynamics is extremely slow at low
temperatures and that for instance equilibration cannot be achieved
any more on laboratory time scales. This is a disadvantage only as
long as one is solely interested in equilibrium quantities, but it
becomes a very fascinating subject for investigation as soon as one
gives up the pretension to explore this regime that is an unnatural
form of existence for glassy materials anyway: they are typically out
of equilibrium. In this sense {\it aging} is just another word for
{\it non-equilibrium} dynamics and means that the outcome of any
experiment that is done with the system under consideration will
depend on the procedures one has applied to it at former times.

Supposed one is interested in the functional form of the
time-dependence of some observable ${\cal O}(t)$, where the point
$t=0$ is chosen deliberately and defines a separation between the
beginning of the observation and the system's former history in the
glassy phase of duration $t_w$, the waiting time or the time for which
the system has {\it aged}. Experimentalists usually apply after this
particular time a change in a field $\hat{\cal O}$ that is conjugate
to the observable ${\cal O}$ so that ${\cal O}(t)$ is essentially the
response of the system to this field change. Instead one might also
measure (e.g.\ in Monte-Carlo simulations) the corresponding
correlation function ${\cal O}(t){\cal O}(t+t_w)$, which would even
contain the same information via the fluctuation dissipation theorem
provided equilibrium conditions would hold. Note that one is not
interested in this information {\it per se}, but only in the characteristic
aging scenario where ${\cal O}(t)$ depends on the whole functional
form $\hat{\cal O}(t_w)$ even for macroscopically large waiting
times $t_w$ of the order of hours, weeks or even years.

To be concrete let us consider the Edwards-Anderson (EA) model of an
Ising spin glass in three dimensions, which (most
probably)\cite{Review,MCreview} has a phase transition into a spin
glass phase at some finite temperature:

\be
H=-\sum_{\langle ij\rangle} J_{ij} S_i S_j\;.
\label{ea}
\ee

The Ising spins $S_i=\pm1$ are put on a simple cubic lattice of linear
dimension $L$ with periodic boundary conditions. The interaction
strengths $J_{ij}$ are quenched random variable obeying a Gaussian
distribution with zero mean and variance one, the spin glass
transition temperature is $T_g\approx0.9$ \cite{Review}. Usually we
are interested in a stochastic, microscopic, single-spin-flip dynamics
that is non-conservative in energy and magnetization, modelling the
coupling of the magnetic moments in real spin glasses to a heat-bath
representing the lattice phonons. Hence we choose the so-called
heat-bath algorithm, which flips single spins with a probability
$w(S_i\to -S_i)={\rm min}\{1,\exp(\Delta E_i/T)\}$, $\Delta E_i$ being
the energy difference $H(S_i)-H(-S_i)$ between the old and the new
configuration. However, the results will not depend significantly on
this choice, as we have checked explicitly.

\begin{figure}
\psfig{file=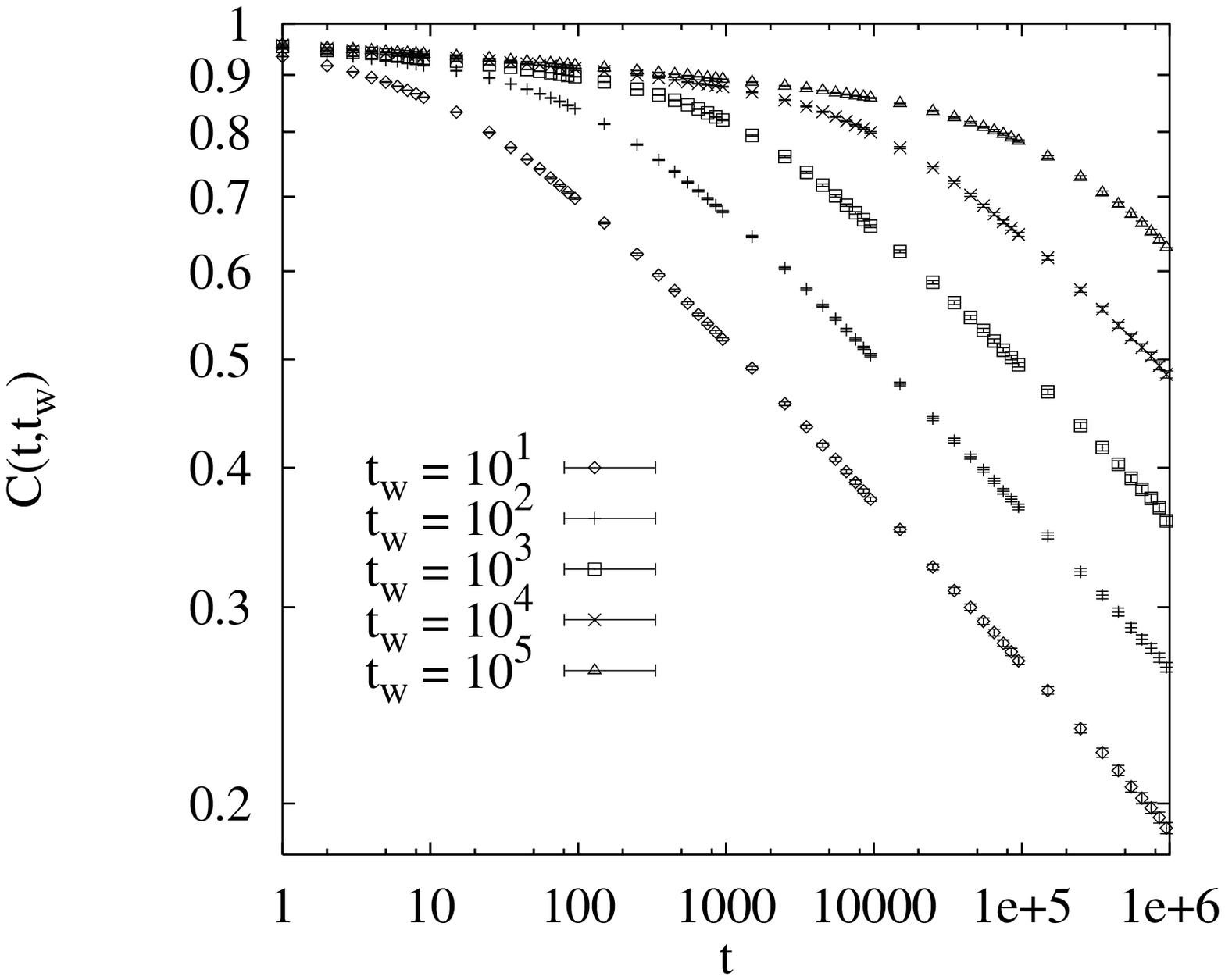,height=18cm}
\caption{Autocorrelation function $C(t,t_w)$ for the EA-spin glass model
  in {\bf three} dimensions as a function of time $t$ for $t_w=5^n$
  ($n=1,\ldots,8$) at $T=0.6$. The system size is $L=24$ and the
  disorder average was performed over 256 samples.}
\label{fig1}
\end{figure}

The quantity that convincingly demonstrates the aging phenomena and
which is best accessible for a {\it quantitative} analysis in
numerical simulations is the spin autocorrelation function

\be
  C(t,t_w)=\frac{1}{N}\sum_i[\langle S_i(t+t_w)S_i(t_w)\rangle ]_{av}\;,
\label{autocorr}
\ee

where for instance $\underline{S}(t_w)$ is the configuration of the
system after the waiting time $t_w$ and time is measured in number of
Monte-Carlo sweeps through the whole lattice. The angular brackets
indicate the average over different initial conditions and
$[\cdots]_{\rm av}$ mean the disorder average. Note that as mentioned
before in experiments usually the corresponding response function,
i.e.\ the thermo-remanent magnetization decay after a field change at
time $t_w$ is measured. In fig.\ \ref{fig1} the result for one
particular temperature (in the spin glass phase) is shown in a log-log
plot. Several characteristic features can be read off immediately,
like the a crossover from a slow {\it quasi-equilibrium} decay for
$t\ll t_w$ to a faster {\it non-equilibrium} decay for $t\gg t_w$ and
the functional form of these decays being algebraic rather than
logarithmic, which can be subsumed in the scaling formula
\cite{age3dpm,age3d}

\be
  C(t,t_w) = t^{-x(T)}\Phi_T(t/t_w)\;,
\label{scale}
\ee

with $\Phi_T(y)=c_T$ for $y=0$ and $\Phi_T(y)\propto
y^{x(T)-\lambda(T)}$ for $y\rightarrow\infty$. The most important
observation is the $t/t_w$ scaling behavior, in contrast to e.g.\ the
prediction $C(t,t_w)\sim (\ln t)^{-\theta/\psi} \tilde{\Phi}\{
\ln(t/\tau)/\ln (t_w/\tau)\}$ by the droplet theory \cite{FiHu}.
Nevertheless we think that the domain growth or coarsening picture of
the latter theory is appropriate: for the length and time scales under
consideration only a basic scaling assumption has to be modified in
order to be consistent with the result (\ref{scale}), see
\cite{MCreview,age3dpm,age3d} for details.

\begin{figure}
\psfig{file=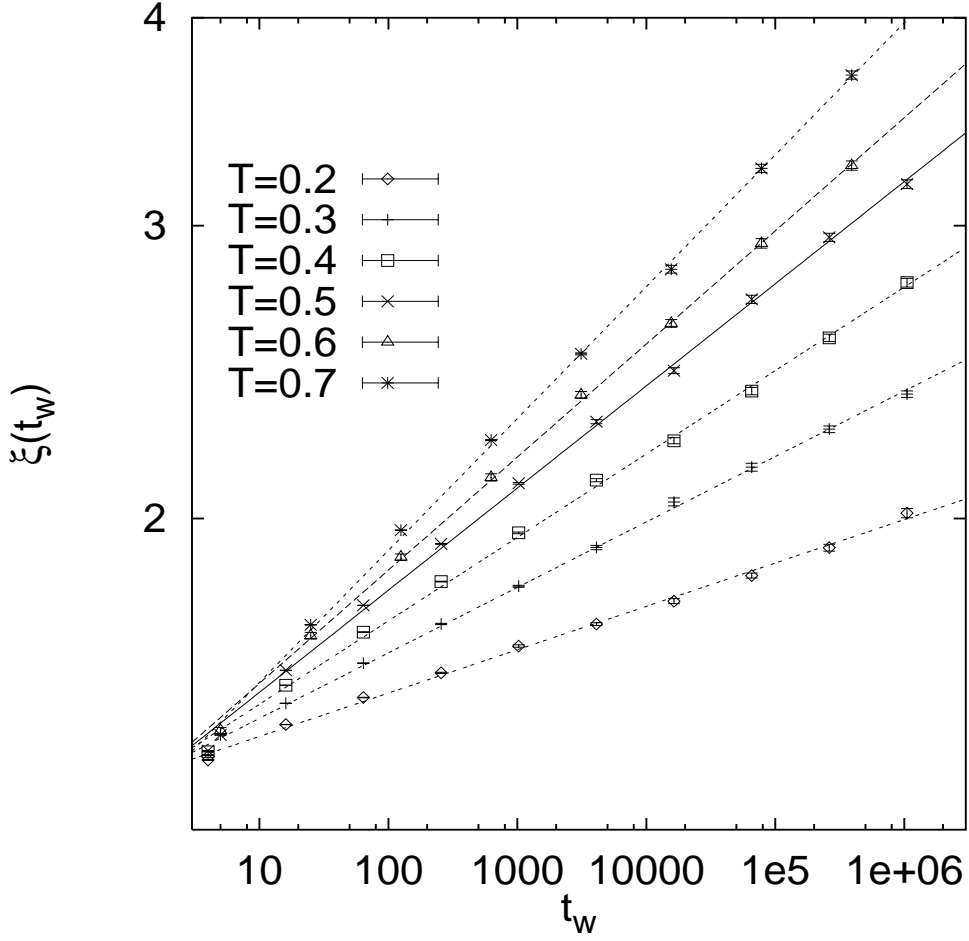,height=18cm}
\caption{Waiting time dependence of the correlation length $\xi(t_w)$
  in three dimensions for different temperatures. The straight lines
  are least square fits to an algebraic growth law $\xi(t_w)\sim
  t_w^{\alpha(T)}$ with the exponent $\alpha(T)$ varying approximately
  linear between $\alpha(T=0.2)=0.026$ and $\alpha(T=0.7)=0.081$.}
\label{fig2}
\end{figure}

The domain growth taking place during the waiting time $t_w$ can be
studied in a straightforward manner within Monte-Carlo simulations by
calculating the spatial correlation function $G(r,t_w)=[\langle
S_i(t_w)S_{i+r}(t_w) \rangle^2]_{av}$. From this one can extract the
correlation length $\xi(t_w)$ either with the help of the expected
scaling form $G(r,t_w)\sim\tilde{g}(r/\xi(t_w))$ or via the integral
$\xi(t_w)=2\int dr\,G(r,t_w)$. In fig.\ \ref{fig2} the waiting time dependence
of this correlation length for a particular temperature is shown in a
log-log plot. It turns out \cite{age3d} that a fit of the waiting time
dependence of this correlation length to an algebraic growth law
$\xi(t_w)\sim t_w^{\alpha(T)}$ works very well \cite{age3d}, which is
consistent with the asymptotic algebraic decay of the autocorrelation
function $C(t,t_w)$ and its $t/t_w$-scaling behavior.

In addition to waiting time experiments described above also other
procedures (in terms of the function $\hat{\cal O}(t_w)$ mentioned in
the first paragraph of this section) have been applied experimentally.
For instance so-called temperature cycling experiments, which consist
of two temperature changes during the time in which the material is
aged in the spin glass phase: either a short heat pulse is applied to
the spin glass during the waiting time after which the relaxation of
e.g.\ the thermo-remanent magnetization is measured, or a short
negative temperature cycle is performed, which is the same as a heat
pulse but with a negative temperature shift during the pulse. It has
been pointed out that this kind of experiments can discriminate
between the droplet picture \cite{FiHu} and the hierarchical picture
\cite{Ledermann}. The interpretation of the experimental situation is
still controversial \cite{Vincent,Mattson} and as long as numerical
simulations follow exactly the lines of the experiments the outcome is
pretty similar \cite{tcycle,overlap}, meaning inconclusive. However,
in numerical studies one has a much broader spectrum of quantities
that can be analyzed and in \cite{age3d} the overlap-correlation
function $[\langle S_i S_{i+r}\rangle_T \langle S_i
S_{i+r}\rangle_{T\pm\Delta T}]_{\rm av}$ was calculated explicitely.
Here no indication of the existence of an overlap-length, which is one
of the underlying concepts of the droplet theory, could be found.

Finally it should be noted that the microscopic theory for the
off-equilibrium dynamics of mean field models of spin glasses has been
pushed forward recently \cite{skage}.  The mathematical difficulties
for an analytically exact solution of the dynamical off-equilibrium
mean field equations for e.g.\ the SK-model come from the lack of the
fluctuation-dissipation theorem (FDT) that relates autocorrelation and
response function. The new approaches circumvent this by considering a
so-called fluctuation-dissipation ratio defined via

\be
x(t,t')={r(t,t')\over \beta\partial C(t,t')/\partial t'}
\label{fdtratio}
\ee

and postulating a particular set of properties for this function
$x(t,t')$ in various asymptotic limits, essentially setting up an
"ultrametric" for timescales. In this way Parisi's static, equilibrium
(!) order parameter function $q(x)$ finds its counter-part in
off-equilibrium dynamics. This scenario has been checked in
numerically in three dimensions \cite{fdt}. Indeed a nontrivial
function $x(q)$ was found (the validity of the FDT-theorem would imply
simply $x(q)=1$) and it seems that these concept might also be
applicable in finite dimensions.

\section{Interrupted Aging}

Two-dimensional spin glass models do not have a spin glass transition
at a finite temperature. This means that the equilibrium correlation
length $\xi_{\rm eq}=\lim_{t_w\to\infty}\xi(t_w)$ (where $\xi(t_w)$ is defined
in the last section) stays finite if $T>0$. Moreover there is a
finite, but very large, equilibration time $\tau_{\rm eq}$
characterized by $\xi(t_w)\sim\xi_{\rm eq}$ for $t_w\geq\tau_{\rm
eq}$.  In terms of a coarsening picture this means that after a
temperature quench domains will steadily grow for a some time and
aging in the sense described above will persist. However, as soon as
the waiting time reaches the order of $\tau_{\rm eq}$ the system is
equilibrated and aging is {\it interrupted}.

Monte-Carlo simulations of the two-dimensional EA-model of an Ising
spin glass (identical to (\ref{ea}) but on a square lattice) indeed
confirm this picture of interrupted aging \cite{age2d}. In fig.\ \ref{fig3}
the autocorrelation function (\ref{autocorr}) is shown for different
temperatures. Here it is nicely demonstrated that for higher
temperatures the system reaches equilibrium and the different curves
for $C(t,t_w)$ collapse onto a single curve $C_{\rm eq}(t)$ for waiting
times large than the equilibration time $\tau_{\rm eq}$.  For lower
temperatures the figure becomes indistinguishable from the three
dimensional case, fig.\ \ref{fig1}, which is a consequence of the fact that
the equilibration cannot take place any more on time scales that are
accessible. This is also the situation that is reported in experiments
on two-dimensional spin glasses \cite{exp2d}. Moreover, it can be
shown \cite{age2d} that for these temperatures the correlation length
grows algebraically with a temperature dependent exponent, similar to
the three-dimensional case. At higher temperatures the domain growth
saturates at some finite correlation length as expected.

Up to now only frustrated systems were considered, however,
glassy dynamics and aging also occurs in non-frustrated systems. We studied
the non-equilibrium dynamics and domain growth in the
random Ising chain \cite{age1d}

\be
H=-\sum_{i=1}^L J_i S_i S_{i+1}\;,
\ee

with random ferromagnetic bonds $J_i$ (note that the distribution does
not need to be confined to positive couplings since their sign can be
removed by a simple gauge transformation). Although this system does not
have any frustration a typical (interrupted) aging scenario at low
temperatures can be observed.  A broad distribution of (free) energy
barriers results in a very slow domain growth results. The latter can
easily be studied in this context, since domains are simply
ferromagnetically ordered segments of the chain.

\section{Non-conventional aging}

 From what has been said in the last section in might occur that only a
very few ingrediences might be necessary in order to observe aging at
least over some time scale. Indeed aging becomes manifest already in
the ferromagnetic Ising chain $H=-J\sum_i S_i S_{i+1}$ at zero
temperature \cite{age1d}, where the dynamics is simply described by
kink diffusion and annihilation. In particular the autocorrelation
function $C(t,t_w)$ scales with $t/t_w$ as observed in many other
models with and without disorder or frustration, too.  However, this
is not the only possible scenario one can think of and some systems
show deviations from this $t/t_w$ scaling, which we call {\it
  non-conventional aging}. For instance it is also possible that after
the temperature quench the system gets trapped in narrow (free) energy
minima of a particular depth in such a way that contiuous domain
growth (or steady equilibration) is not possible any more. In this
way, very reminiscent of the structural glass transition, the system
is frozen into an amorphous metastable state for an astronomically
long time.

To illustrate this sort of behavior we studied a very simple spin model
without any frustration or disorder that possesses many glass like
features at low temperatures. It is defined by the following $p$-spin
interaction Hamiltonian for Ising spins on a chain \cite{pspin}

\begin{equation}
H=-J\sum_{i=1}^L \,S_i S_{i+1} \cdots S_{i+p-1}\;.
\label{hampspin}
\end{equation}

For simplicity we consider the case $p=3$.
The groundstate of this system has a 4--fold degeneracy (in general
$2^{p-1}$). Introducing a local energy-variable $\tau_i = S_{i-1}S_i S_{i+1}$
all groundstate configurations are described by $\tau_i=+1$ for all
sites $i$. Consider a configuration in $\tau$ and $S$ variables

\begin{equation}
\begin{array}{cccc}
(\underline{\tau}) & \cdots +++++++++++&-&+++++++++++++++ \cdots\\
(\underline{S})    & \cdots +++++++++++&+&--+--+--+--+--+ \cdots\\
                  & & i &
\end{array}
\label{config}
\end{equation}

which consists of two domains, both being in a minimum energy
configuration, separated by a domain wall located at site $i$.  It
costs an energy of $2J$ to move the domain wall at position $i$ to the
right or left, thus the system is frozen into such a metastable
configuration for a time $t_{\rm freeze}\approx\exp(2J/T)$.  Moreover,
it can be shown that all configurations of the type (\ref{config})
consisting (expressed in $\tau$-variables) of strings of arbitrary
length $l\ge2$ with $\tau=+1$ separated by isolated sites with
$\tau=-1$ are indeed metastable. In a chain of length $L$ an
exponentially large number $n_{\rm stable}\sim L^{1.4655}$ of these
configurations exist. Starting with a random initial state the
sequential update procedure at zero temperature will drive the system
into one of this exponentially large number of metastable states
within only two sweeps through the whole chain.  Thus after $2t_0$,
where $t_0$ is the microscopic time scale, the system will be frozen
for a time $t_{\rm freeze}=t_0\exp(2J/T)$.  This can be seen for
instance by looking at the waiting time depends of the average domain
size, which enters a plateau at a small value (which can be calculated
exactly to be 5.775 lattice spacings \cite{pspin}) after two time
steps that extends to infinity for $T\to0$ as is shown in fig.\
\ref{fig4}.

As mentioned before this is an extremely simplified model with one
single characteristic energy barrier that prevents the the system from
relaxing into its equilibrium configuration for an exponentially large
time. By allowing the interaction strengths to vary spatially and
arranging them for instance in a hierarchical way one generates
spatially varying energy barriers and a broad distribution of
exponentially large trapping times. On the other hand it is possible
to generalize this model to higher dimensions. In this case even a
true equilibrium phase transition is possible and a closer contact to
the physics of the structural glass transition might become feasible.

\section{Summary}

As we have seen the non-equilibrium dynamics of strongly disordered
systems is a very fascinating subject. Experimentally as well as
numerically one is confined to rather restricted time and length
scales for which a complete theory is still missing (although some
progress has been announced \cite{Hilhorst}). Aging does not only
occur in frustrated systems, as has been demonstrated in the random
bond Ising chain, and can even be observed in systems without any
disorder.  Most systems show an aging behavior in which spin
autocorrelations scale with $t/t_w$ and the domain growth depends
algebraically on the waiting time $t_w$.  However also other scenarios
are possible and models are currently under investigation that are
very reminiscent of the structural glass transition.

\section*{Acknowledgements}
I would like to express my special thanks towards my
collaborators J.\ Kisker, L.\ Santen, B.\ Steckemetz, M.\
Schreckenberg and S.\ Franz. I acknowledge the allocation of computing
time on the Intel Paragon XP/S 10 of the HLRZ at the Forschungszentrum
J\"ulich and on the Parsytec--GCel1024 of the Center of Parallel
Computing (ZPR) in K\"oln. This work was performed within the SFB 341
K\"oln--Aachen--J\"ulich.

\begin{figure}
\caption{Autocorrelation function $C(t,t_w)$ for the EA-spin glass model
  in {\bf two} dimensions as a function of time $t$ for $t_w=5^n$
  ($n=1,\ldots,8$) at $T=1.0$ and $0.8$, ($n=2,\ldots,8$) at $0.6$ and
  $0.2$. The system size is $L=100$ and the disorder average was
  performed over 256 samples. The errorbars are smaller than the
  symbols.}
\label{fig3}
\end{figure}
\psfig{file=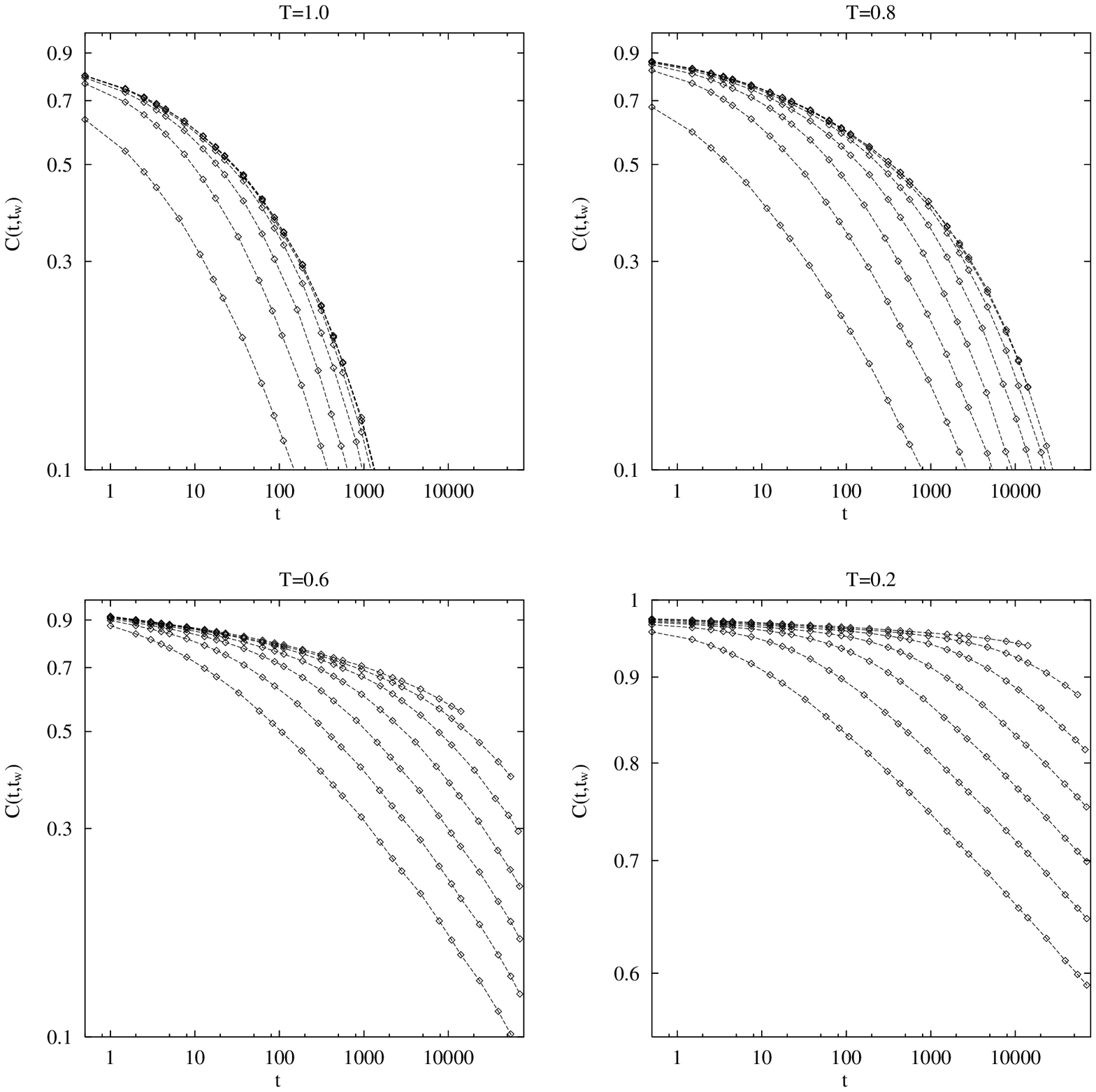,height=18cm}

\begin{figure}
\psfig{file=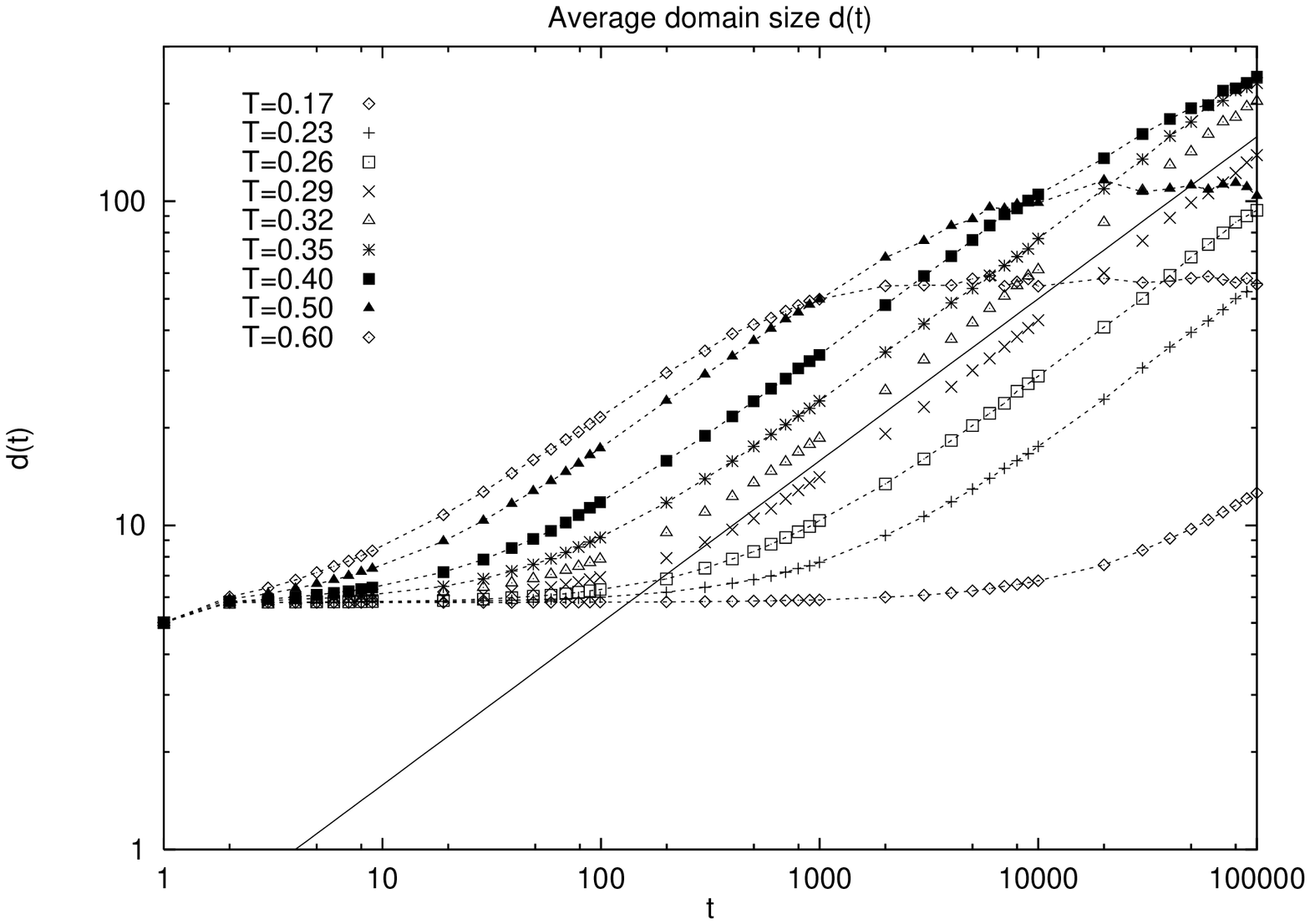,height=18cm}
\caption{Average domain size in dependence of the waiting time $t$
  of the Ising chain with 3-spin interactions (\ref{hampspin}) for
  various temperatures calculated via Monte-Carlo simulation of a
  system with $10^6$ spins. The intermediate growth (between melting
  of the frozen domains and final saturation by equilibration) can be
  fitted nicely to $d(t)\sim t^{1/2}$ (solid line).}
\label{fig4}
\end{figure}


\begin{references}

\bibitem{Review}
        K.~Binder and A.~P.~Young, Rev.\ Mod.\ Phys.\ {\bf 58}, 801 (1986);
        D.\ Chowdhury, {\it Spin glasses and other frustrated systems},
        World Scientific, Singapore (1986).

\bibitem{MCreview}
        For a recent review see H. Rieger, {\em Monte Carlo studies of
        spin glasses and random field systems} in
        Annual Reviews of Computational Physics II, p.\ 295 (1995).

\bibitem{glasses}
        J.\ J\"ackle, Rep.\ Prog.\ Phys.\ {\bf 49}, 171 (1986);
        W.\ G\"otze and L.\ Sj\"ogren, Rep.\ Prog.\ Phys.\
        {\bf 55}, 241 (1992);
        for a series of review papers on glasses see:
        Science {\bf 267} 1924--1953 (1995).

\bibitem{manifolds}
        B.\ Derrida and H.\ Spohn, J.\ Stat.\ Phys.\ {\bf 51}, 817 (1988);
        M\'ezard and G.\ Parisi, J.\ Phys.\ A {\bf 23}, L1229 (1990);
        H.\ Kinzelbach and H.\ Horner,
        J.\ Phys.\ I France {\bf 3}, 1901 (1993).

\bibitem{protein}
        J.\ B.\ Bryngelson and P.\ G.\ Wolynes,
        Proc.\ Nat.\ Acad.\ Sci.\ USA {\bf 82}, 3670 (1985);
        E.\ I.\ Shaknovich and A.\ M.\ Gutin,
        Europhys.\ Lett.\ {\bf 8}, 327 (1989).
        I.\ E.\ T.\ Iben et al., Phys.\ Rev.\ Lett.\ {\bf 62}, 1916  (1989);

\bibitem{super}
        S.\ John and T.\ C.\ Lubensky, Phys.\ Rev.\ B {\bf 34}, 4815 (1986);
        D.\ S.\ Fisher, M.\ P.\ A.\ Fisher and D.\ A.\ Huse,
        Phys.\ Rev.\ B {\bf 43}, 130 (1991).

\bibitem{CDW}
        G.\ Kriza and G.\ and G.\ Mih\'aly,
        Phys.\ Rev.\ Lett.\ {\bf 56}, 2529 (1986);
        P.\ B.\ Littlewood and R.\ Rammal,
        Phys.\ Rev.\ B {\bf 38}, 2675 (1988).

\bibitem{biology}
        For an overview see R.\ Livi et al.\ ed.\ {\it Chaos and Complexity},
        World Scientific, Singapore (1988);
        D.\ L.\ Stein ed.\ {\it Spin glasses and biology}, World Scientific,
        Singapore (1992) and references therein.

\bibitem{Lundgren}
        L.~Lundgren, R.~Svedlindh, P.~Nordblad and O.~Beckman,
        Phys.\ Rev.\ Lett.\ {\bf 51}, 911 (1983).

\bibitem{Ledermann}
        M.~Ledermann et al., Phys.\ Rev.\ B {\bf 44}, 7403 (1991).

\bibitem{Vincent}
        E. Vincent, J. Hammann and M. Ocio in
        {\em Recent Progress in Random Magnets} (World
        Scientific, Singapore, 1992), p.\ 207.

\bibitem{Mattson}
        J.~Mattson et al., Phys.\ Rev.\ B {\bf 47}, 14626 (1993).

\bibitem{Struik}
        L.\ C.\ E.\ Struik {\it Physical ageing in amorphous polymers
        and other materials}, Elsevier Scientific Publ.\ Co.\ (1978).

 \bibitem{Rossel}
        C.\ Rossel, Y.\ Maeno and I.\ Morgenstern,
        Phys.\ Rev.\ Lett.\ {\bf 62}, 681 (1989).

\bibitem{Biljakovic}
        K.\ Biljakovic, J.\ C.\ Lasjaunias, P.\ Monceau and F.\ Levy,
        Phys.\ Rev.\ Lett.\ {\bf 62}, 1512 (1989)
        and ibid.\ {\bf 67}, 1902 (1991).

\bibitem{exp2d}
        A.~Schins et al., Phys.\ Rev.\ B {\bf 48}, 16524 (1993).
        J.~Mattson et al., Phys.\ Rev.\ B {\bf 47}, 14626 (1993).

\bibitem{skage}
        L.~Cugliandolo and J.~Kurchan, J.\ Phys.\ A {\bf 27}, 5749 (1994).
        S.\ Franz and M.\ M\'ezard, Physica A {\bf 210}, 43 (1994).

\bibitem{Bouchaud}
        J.~P. Bouchaud, J. Physique I 2 (1992) 1705;
        J.~P. Bouchaud and D.~S. Dean, J. Physique I 5 (1995) 265;
        E.\ Marinari, G.\ Parisi, J.\ Phys.\ A {\bf 26}, L1149 (1993).

\bibitem{FiHu}
        D.~S.~Fisher and D.~A.~Huse,
        Phys.\ Rev.\ B {\bf 38}, 373 (1988) and ibid.\ {\bf 38}, 386 (1988).

\bibitem{KoHi}
        G.~J.~M. Koper and H.~J.~Hilhorst,
        J.\ Phys.\ France {\bf 49}, 429 (1988).

\bibitem{age3dpm}
        H. Rieger, J. Phys. A {\bf 26}, L615 (1993).

\bibitem{age3d}
        J.\ Kisker, L.\ Santen, M.\ Schreckenberg and H.\ Rieger,
        Phys.\ Rev.\ B, in press (1995).

\bibitem{tcycle}
        H. Rieger, J. Physique I {\bf 4}, 883 (1994).

\bibitem{overlap}
        J.~O.~Andersson, J.~Mattson and P.~Svedlindh,
        Phys.\ Rev.\ B {\bf 49}, 1120 (1994).

\bibitem{fdt}
        S. Franz and H. Rieger, J. Stat. Phys. {\bf 79}, 749 (1995).

\bibitem{age2d}
        H. Rieger, B. Steckemetz and M. Schreckenberg,
        Europhys. Lett. {\bf 27}, 485 (1994).

\bibitem{age1d}
        H. Rieger, J. Kisker and M. Schreckenberg,
        Physica A {\bf 210}, 326 (1994).

\bibitem{pspin}
        J. Kisker, H. Rieger and M. Schreckenberg,
        J. Phys. A {\bf 27}, L853 (1994).

\bibitem{Hilhorst}
        H.\ Hilhorst, to be published (1995).


\end{references}
\end{document}